\documentclass[a4paper,11pt]{article}
\usepackage[utf8]{inputenc}
\usepackage{booktabs} 
\usepackage{multirow} 
\usepackage{array}
\usepackage[margin=0.9in]{geometry}
\usepackage{multicol}
\usepackage{mathtools}
\usepackage{enumitem}
\usepackage{setspace}
\normalsize
\usepackage{bm}
\usepackage{float}
\usepackage[framemethod=tikz]{mdframed}
\usepackage{lipsum}
\usepackage[toc,page]{appendix}
\usepackage{graphics}
\usepackage[utf8]{inputenc}
\usepackage[english]{babel}
\usepackage{hyperref}
\usepackage{float}
\usepackage{graphicx}
\usepackage{wrapfig}
\usepackage{pgf} 
\usepackage[utf8]{inputenc}
\usepackage{graphicx}
\usepackage{comment}
\usepackage{lscape} 
\usepackage{framed}
 \usepackage{multirow}
    \usepackage{amsmath}
    \usepackage{subcaption}
	\usepackage{listings}
\usepackage{soul}
    
    \usepackage{graphicx}
    \usepackage{dcolumn}
    \usepackage{bm}
    \usepackage{amsmath} 
\usepackage{colortbl}
\newcolumntype{L}[1]{>{\raggedright\let\newline\\\arraybackslash\hspace{0pt}}m{#1}}
\usepackage[toc,page]{appendix}

\hypersetup{
  colorlinks,
  allcolors=.,
  urlcolor=blue,
}
\usepackage{amsmath}
 
\usepackage{float}
 \usepackage{multirow}
 \usepackage{booktabs} 
\title{Spatial Non-parametric Bayesian Clustered 
Coefficients}
\author{Wala Draidi Areed, Aiden Price, Helen Thompson, Reid Malseed, Kerrie Mengersen}
\date{}

\begin{document}

\maketitle

\begin{abstract}
In the field of population health research, understanding the similarities between geographical areas and quantifying their shared effects on health outcomes is crucial. In this paper, we synthesise a number of existing methods to create a new approach that specifically addresses this goal. The approach is called a Bayesian spatial Dirichlet process clustered heterogeneous regression model. This non-parametric framework allows for inference on the number of clusters and the clustering configurations, while simultaneously estimating the parameters for each cluster. We demonstrate the efficacy of the proposed algorithm using simulated data and further apply it to analyse influential factors affecting children's health development domains in Queensland. The study provides valuable insights into the contributions of regional similarities in education and demographics to health outcomes, aiding targeted interventions and policy design.
\end{abstract}

\flushbottom
\maketitle
%
%


\section*{Introduction}

Collecting data related to places has become easier than ever with modern technology.  To date, the most common methods used to model geographically referenced data are spatial linear regression \cite{cressie2015statistics} and spatial generalized linear regression \cite{diggle1998model}.  However, these models assume that the coefficients of the explanatory variables are constant across space, which can be overly restrictive for large regions where the regression coefficients may vary and potentially cluster spatially. 
Spatially varying coefficient models have been proposed, such as the geographically weighted regression (GWR), which fits a locally weighted regression model at each observation, with a kernel function defining the weight matrix \cite{fotheringham1998geographically}. Spline approaches have been explored to estimate bivariate regression functions \cite{opsomer2008non} and to accommodate irregular domains with complex boundaries or interior gaps \cite{wang2007low}. Other studies, including those by Li et al. \cite{li2021sparse} and Wang et al. \cite{wang2020efficient}, also address this problem over irregular domains. However, all of these methods have a significant limitation: they cannot handle the possibility of a spatially clustered pattern in the regression coefficients.

 Some authors have proposed methods to address this challenge. A recent development by Li et al. \cite{li2019spatial} is the spatially clustered coefficient (SCC) regression, which employs the fused LASSO to automatically detect spatially clustered patterns in the regression coefficients. Ma et al. \cite{ma2020heterogeneous} and Luo et al. \cite{luo2021bayesian} have proposed the spatially clustered coefficient models using Bayesian approaches. Ma et al. \cite{ma2020heterogeneous} identified coefficient clusters based on the Dirichlet process, whereas Luo et al. \cite{luo2021bayesian} used a hierarchical modelling framework with a Bayesian partition prior model from spanning trees of a graph. However, these approaches assume that the values of regression coefficients are constant within each cluster. This constraint may result in overestimation of the number of clusters when the true regression coefficients vary smoothly within each cluster.
 
 Achieving unbiased estimation of regression coefficients and valid inference is crucial in regression analysis. However, spatial heterogeneity or structural instability can result in biased estimators and incorrect inference, making it challenging to identify the correct grouping structure of the regression coefficients. This is exacerbated if the relationships among spatial variables exhibit changes across certain boundaries. Moreover, all of these estimates and inferences are made with some uncertainty. This leads to the need for new methods to identify spatially clustered patterns with uncertainty measures in these relationships. This paper presents the Bayesian approaches for detecting contiguous clusters in regression coefficients.\\
 The selection of the appropriate number of clusters is a crucial aspect of clustering analysis. Most traditional methods require the number of clusters to be specified beforehand, which can limit their applicability in practice. Dirichlet process mixture models (DPMM) have gained popularity in Bayesian statistics as they allow for an unknown number of clusters, increasing the flexibility of clustering analysis. However the DPMM does not account for the spatial information in the clusters. 
 
In this paper, we synthesise two approaches, namely, a Bayesian GWR and a Bayesian spatial DPMM, to create a new method called  the Bayesian spatial Dirichlet process clustered heterogeneous regression model. This method can detect spatially clustered patterns while considering the smoothly varying relationship between the response and the covariates within each group. We used a Bayesian geographically weighted regression algorithm to model the varying coefficients over the geographic regions and incorporated spatial neighbourhood information of regression coefficients. We then combined the regression coefficient and a spatial Dirichlet mixture process to perform the clustering. The approach is demonstrated using simulated data and then applied to a real-world case study on children's development in Queensland, Australia.

 The strength of the proposed clustering method lies in several key features that set it apart from traditional clustering algorithms. Unlike $K$-Means and hierarchical clustering (HC), which lack uncertainty measures, the proposed method provides clusters with associated uncertainty measures, enhancing the interpretability and making them more valuable for decision-making and analysis. Additionally, the proposed method incorporates spatial neighbourhood information, which a crucial aspect when dealing with geospatial data. This ensures that the resulting clusters not only reflect data similarity but also account for spatial heterogeneity, a critical consideration in practical applications. Furthermore, the Bayesian approach proposed in this paper enhances its robustness, outperforming traditional algorithms like $K$-Means, HC, and partition around medoids (PAM). This Bayesian framework allows for better handling of outliers and uncertainties in the data by incorporating prior information. This adaptability is particularly beneficial in scenarios where data quality varies or is incomplete.

\section*{Results}
The proposed method, described in detail in the Methods section below, is evaluated through a simulation study and applied to a real-world case study. These applications demonstrate the effectiveness of the approach in simultaneously estimating and clustering spatially varying regression coefficients, with associated measures of uncertainty.
\section*{Simulation Study}
 The simulation was structured based on the Georgia dataset with 159 regions introduced by Ma \cite{ma2020heterogeneous}, where spatial sampling locations represented geographical positions for data collection. Specifically, we used centroids of geographical areas as the sampling locations. 

For the simulation, six covariates ($X_1$ to $X_6$) were introduced as independent variables, each representing distinct features or characteristics at each sampling location. To incorporate spatial autocorrelation, we generated the covariates using multivariate normal distributions with spatial weight matrices derived from the distance matrix and parameter bandwidth.

The response variable ($Y$) in the simulation was generated using the GWR model \cite{fotheringham1998geographically}:
\begin{equation}
y(s) = \beta_{0}(u(s), v(s)) + \sum_{k=1}^{K} \beta_{k}(u(s), v(s)) \cdot X_{k}(s) + \varepsilon(s)
\end{equation}

 It is noteworthy that the true parameters ($\beta_1$ to $\beta_6$) of the GWR model varied spatially, implying that they differed across sampling locations based on the spatial weight matrices. This spatial variation allowed us to capture spatially dependent effects in the simulation \cite{sugasawa2022adaptively}.

\begin{figure}
    \centering
    \includegraphics{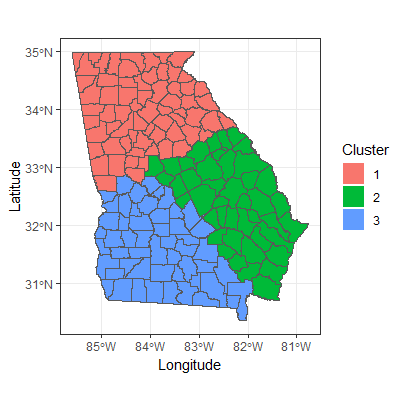}
    \caption{Regional cluster assignment for Georgia counties used for simulation study.}
    \label{fig:enter-labelw}
\end{figure}

To create distinct spatial patterns in the data, we visually partitioned the counties of Georgia into three large regions based on the spatial coordinates of centroids, defining true clustering settings. This approach enabled us to incorporate spatial autocorrelation, spatial variability, and true clustering effects in the simulated data. Figure \ref{fig:enter-labelw} visualizes partition of the counties into three large regions with sizes, 51, 49 and 59 areas.
 The code for the proposed algorithm can be found in the first author's GitHub \url{https://github.com/waladraidi/Spatial-stick-breaking-BGWR}. 

\begin{figure}
    \centering
\includegraphics[scale=0.5]{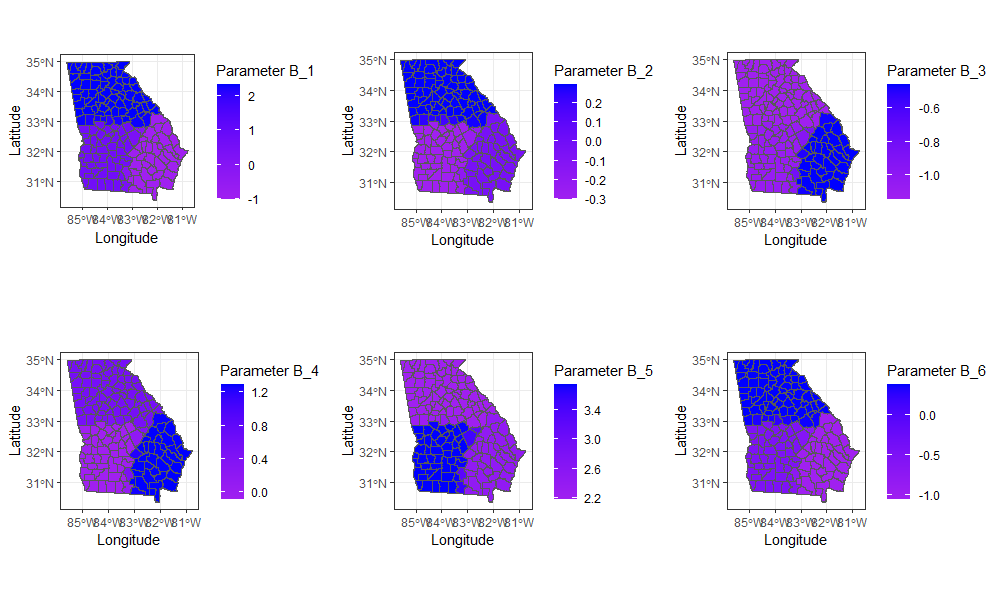}
    \caption{The spatial distribution of the posterior mean for the parameters obtained from the proposed model.}
    \label{fig:enter-labelqq}
\end{figure}

The simulation was repeated 100 times. Figure \ref{fig:enter-labelqq} illustrates the spatial distribution of the posterior mean parameter coefficients for each location over the 100 replicates. This figure showcases the diverse spatial patterns and disparities in these parameter coefficients across the study area, providing insights into the geographical variation.

The performance of these posterior estimates was evaluated by mean absolute bias (MAB), mean standard deviation (MSD), mean of mean squared error (MMSE), as follows:

\begin{equation}
MAB = \frac{1}{159}\sum_{\ell=1}^{159} \frac{1}{100}\sum_{r=1}^{100} |\hat{\beta}_{\ell, m, r} - \beta_{\ell, m}|, \label{eq:19}
\end{equation}

\begin{equation}
MSD = \frac{1}{159}\sum_{\ell=1}^{159} \sqrt{\frac{1}{99}\sum_{r=1}^{100} (\hat{\beta}_{\ell, m, r} - \bar{\hat{\beta}}_{\ell, m})^2}, \label{eq:20}
\end{equation}

\begin{equation}
MMSE = \frac{1}{159}\sum_{\ell=1}^{159} \frac{1}{100}\sum_{r=1}^{100} (\hat{\beta}_{\ell, m, r} - \beta_{\ell, m})^2, \label{eq:21}
\end{equation}

\noindent where $\bar{\hat{\beta}}_{\ell, m}$, is the average parameter estimate and has been calculated by the average of $\hat{\beta}_{\ell,m}$ ($\ell = 1, \ldots, 159; \quad m = 1, \ldots, 6$) in 100 simulations, and $\hat{\beta}_{\ell, m, r}$ denotes the posterior estimate for the $m$-th coefficient of county $\ell$ in the $r$-th replicate.
In each replicate, the MCMC chain length is set to be 10,000, and the first 2000 samples are discarded as burn-in. Therefore, we have 8000 samples for posterior inference. Table \ref{mytab32} reports the the three performance measures in equations \eqref{eq:19}–\eqref{eq:21} for the simulated data. The parameter estimates are very close to the true underlying values and have a small MAB, MSD, and MMSE.
\begin{table}[ht!]
\centering
\caption{The performance of parameter estimates from the proposed model, where MAE: mean absolute error, MSE: mean squared error, and MMSE: mean of mean squared error.}
\begin{tabular}{llll}
\hline
parameter & MAE  & MSE  & MMSE \\ \hline
$\hat{\beta_1 }$ & 0.84 & 1.06 & 0.23 \\
$\hat{\beta_2 }$ & 0.32 & 0.42 & 1.35 \\
$\hat{\beta_3 }$ & 0.37 & 0.26 & 1.98 \\
$\hat{\beta_4 }$& 0.37 & 0.42 & 1.19 \\
$\hat{\beta_5 }$& 1.37 & 2.72 & 0.43 \\
$\hat{\beta_6 }$ & 0.94 & 1.21 & 0.55 \\ \hline
\end{tabular}
\label{mytab32}
\end{table}

\begin{figure}[ht!]
     \centering
     \includegraphics[scale=0.5]{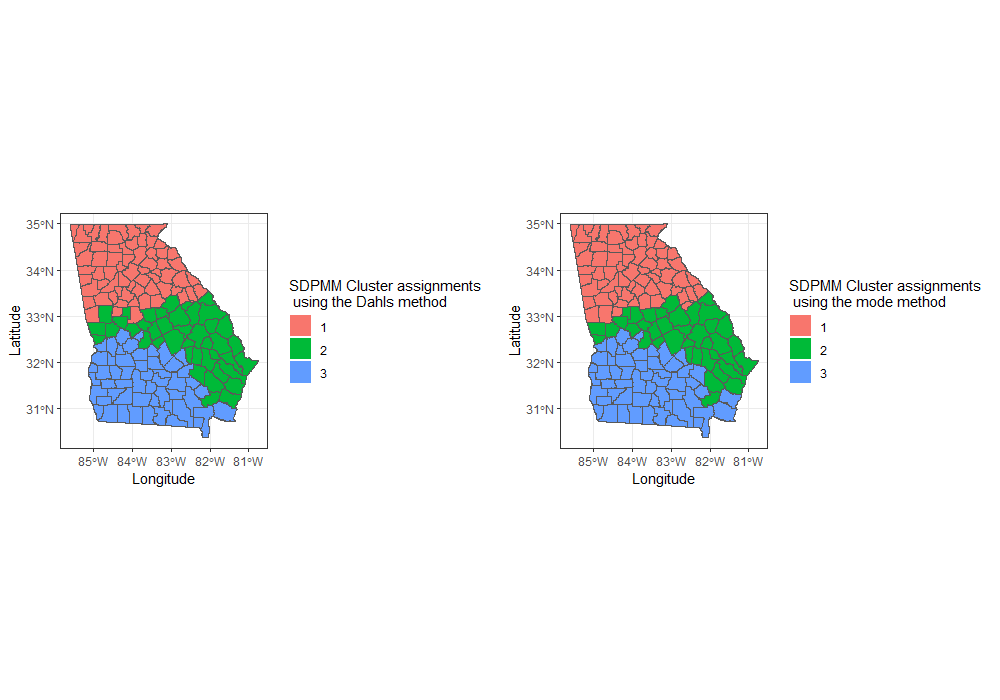}
     \caption{LHS) Cluster assignment for Georgia counties using Dahl's method from the proposed algorithm. RHS) The cluster assignment obtained from the proposed algorithm using the mode method.}
     \label{fig:enter-labelww}
 \end{figure}

Three distinct clusters were found within the 159 regions. The spatial layout of these clusters is visualized in Figure \ref{fig:enter-labelww},  where two cluster configurations are described in the later section on cluster configurations.  Notably, when examining Figure \ref{fig:enter-labelww}, it is clear that the cluster assignments derived from Dahl's and mode allocation approaches (see Methods section below) exhibit a high degree of similarity. The corresponding 
parameter estimates are shown in Table \ref{table333}.

\begin{table}[ht!]
\centering
\caption{Parameter estimates and their 95\% highest posterior density (HPD) intervals for the three
clusters identified.}
\begin{tabular}{cccc}
\toprule
Cluster & 1 & 2 & 3 \\
\midrule
$\hat{\beta}_1$ & -1.0 & 0.62 & 2.31 \\
 & (-1.01, -0.92) & (-0.03, 0.68) & (2.06, 2.32) \\
$\hat{\beta}_2$ & -0.04 & -0.29 & 0.29 \\
 & (-0.06, -0.04) & (-0.30, -0.19) & (0.19, 0.30) \\
$\hat{\beta}_3$ & -0.46 & -1.06 & -1.14 \\
 & (-0.49, -0.46) & (-1.07, -0.82) & (-1.14, -1.13) \\
$\hat{\beta}_4$ & 1.29 & -0.06 & 0.53 \\
 & (1.22, 1.29) & (-0.09, 0.47) & (0.44, 0.54) \\
$\hat{\beta}_5$ & 2.46 & 3.71 & 2.19 \\
 & (2.46, 2.53) & (3.24, 3.74) & (2.18, 2.42) \\
$\hat{\beta}_6$ & -1.05 & -0.51 & 0.37 \\
& (-1.05, -1.02) & (-0.73, -0.49) & (0.24, 0.37) \\
\bottomrule
\end{tabular}
\label{table333}
\end{table}

Since the kernel type, bandwidth prior and the number of knots play a crucial role in the spatial stick-breaking construction of the the proposed mode for the DPMM, the priors and kernel functions from Table \ref{table_true} (Method section) were utilised to test the accuracy of the proposed model. We explored different options to determine the optimal fit for the data using the Watanabe-Akaike onformation criterion (WAIC) \cite{watanabe2013widely}, and we performed a sensitivity analysis on the proposed model with respect to number of knots. The results are summarized in Table \ref{mytab3}.
\begin{table}[ht!]
\centering
\caption{Sensitivity analysis for the number of knots in the spatial stick-breaking with the squared exponential kernel.}
\begin{tabular}{cc}
\toprule
Number of knots & WAIC \\
\midrule
9     &   117044.3  \\
14    &    117066.4  \\
19    &     117051.6 \\
29    &  117055.4\\
\bottomrule
\end{tabular}
\label{mytab3}
\end{table}

According to the Table \ref{mytab3}, the optimal number of knots for the simulated data as 9, as evidenced by the lowest WAIC value. In Table \ref{tab:my-table5}, we present a sensitivity analysis for our proposed algorithm, discussing its performance under various bandwidth priors and kernel functions. This table categorizes the results under two primary kernel types: uniform and squared exponential. 
\begin{table}[ht!]
\centering
\caption{Sensitivity analysis for the proposed algorithm with different bandwidth priors and kernel functions with the number of knots is 9.}
\begin{tabular}{lllll}
\hline
                                 & bandwidth prior & RI   &  Summary                                                          & WAIC     \\ \hline
\multirow{2}{*}{Uniform kernel} &   $\varepsilon_{1i}, \varepsilon_{2i} \equiv \lambda$                  & 0.86 & \begin{tabular}[c]{@{}l@{}}C1=56,\\ C2=70,\\ C3=33\end{tabular}          & 117067.4 \\ \cline{2-5} 
                                 &     $\varepsilon_{1i}, \varepsilon_{2i} \equiv Exp(\lambda)$            & 0.87 & \begin{tabular}[c]{@{}l@{}}C1=51,\\ C2=42,\\ C3=10\end{tabular}          & 117058.4 \\ \hline
\multirow{2}{*}{Exp kernel}    &       $\varepsilon_{1i}, \varepsilon_{2i} \equiv \frac{\lambda^2}{2}$              & 0.88 & \begin{tabular}[c]{@{}l@{}}C1=54, \\  C2=66,\\  C3=39\end{tabular}       & 117044.3 \\ \cline{2-5} 
                                 &      $\varepsilon_{1i}, \varepsilon_{2i} \sim \text{Inverse Gamma }(1.5, \lambda^2/2)$           & 0.84 & \begin{tabular}[c]{@{}l@{}}C1=21,\\ C2=55,\\ C3=39,\\ C4=44\end{tabular} & 126134.4 \\ \cline{1-5} 
\end{tabular}
\label{tab:my-table5}
\end{table}
For each kernel type, two bandwidth priors were evaluated. Based on the WAIC values, the squared exponential kernel with a bandwidth prior of $\varepsilon_{1i}, \varepsilon_{2i} \equiv \frac{\lambda^2}{2}$ emerged as the most effective. Importantly, our algorithm not only demonstrated the stability of clusters when compared to two established methods but also managed to accurately assign clusters with an accuracy of 0.87. In the simulated dataset, our method effectively identified three distinct clusters.
\section*{Real Data analysis}
\subsection*{Case Study}
Children's Health Queensland (CHQ) has developed the CHQ Population Health Dashboard, a remarkable resource providing data on health outcomes and socio-demographic factors for a one-year period (2018-2019) across 528 small areas (Statistical area level 2 (SA2) in Queensland, Australia. The dashboard presents over 40 variables in a user-friendly format, with a focus on health outcomes, particularly vulnerability indicators measuring children's developmental vulnerability across five Australian Early Development Census (AEDC) domains. These domains include physical health, social competence, emotional maturity, language and cognitive skills, and communication skills with general knowledge.

The AEDC also includes two additional domain indicators: vulnerable on one or more domains (Vuln 1) and vulnerable on two or more domains (Vuln 2). Socio-demographic factors, including Socio-Economic Indexes for Areas (SEIFA) score, preschool attendance, and remoteness factors are also incorporated, offering insights into potential links to health outcomes. The SEIFA score summarizes socio-economic conditions in an area, while remoteness factors categorize regions into cities, regional, and remote areas.

In the field of population health, publicly available data are often grouped according to geographical regions, such as the statistical areas (SA) defined in the Australian Statistical Geography Standard (ASGS). These areas, called SA1, SA2, SA3, and SA4, range respectively from the smallest to the largest defined geographical regions. Due to privacy and confidentiality concerns, personal-level information is typically not released. Therefore, in this paper, we focus on group-level data, and in the case study, we use data that have been aggregated at the SA2 level \cite{buchin2008clusters}.

Data for the analysis are sourced from the 2018 AEDC, and focus on the proportion of vulnerable children in each SA2. Some missing data is handled through imputation using neighboring SA2s, with two islands having no contiguous neighbors excluded from the analysis. The study utilises the remaining data from 526 SA2 areas to conduct the analysis.

Our study uses the proposed methodology to analyse the influential factors affecting the development of children who are vulnerable in one or more domains (Vuln 1) in the Queensland SA2 regions. Data were found on the Australian Bureau of Statistics (ABS) official website and the AEDC. For each SA2 region, we considered several dependent variables, including the proportion of attendance at preschool, the remoteness factor which is converted using the one hot coding to three variables including zero and one and the index of relative socio-economic disadvantage (IRSD) factor which is considered continuous in this case study. Before fitting the model, we scaled the variables using the logarithm. As a result, all the models are fitted without an intercept term.


\subsection*{Spatial Cluster Inferences}
\begin{figure}[ht!]
     \centering
     \includegraphics[scale=0.4]{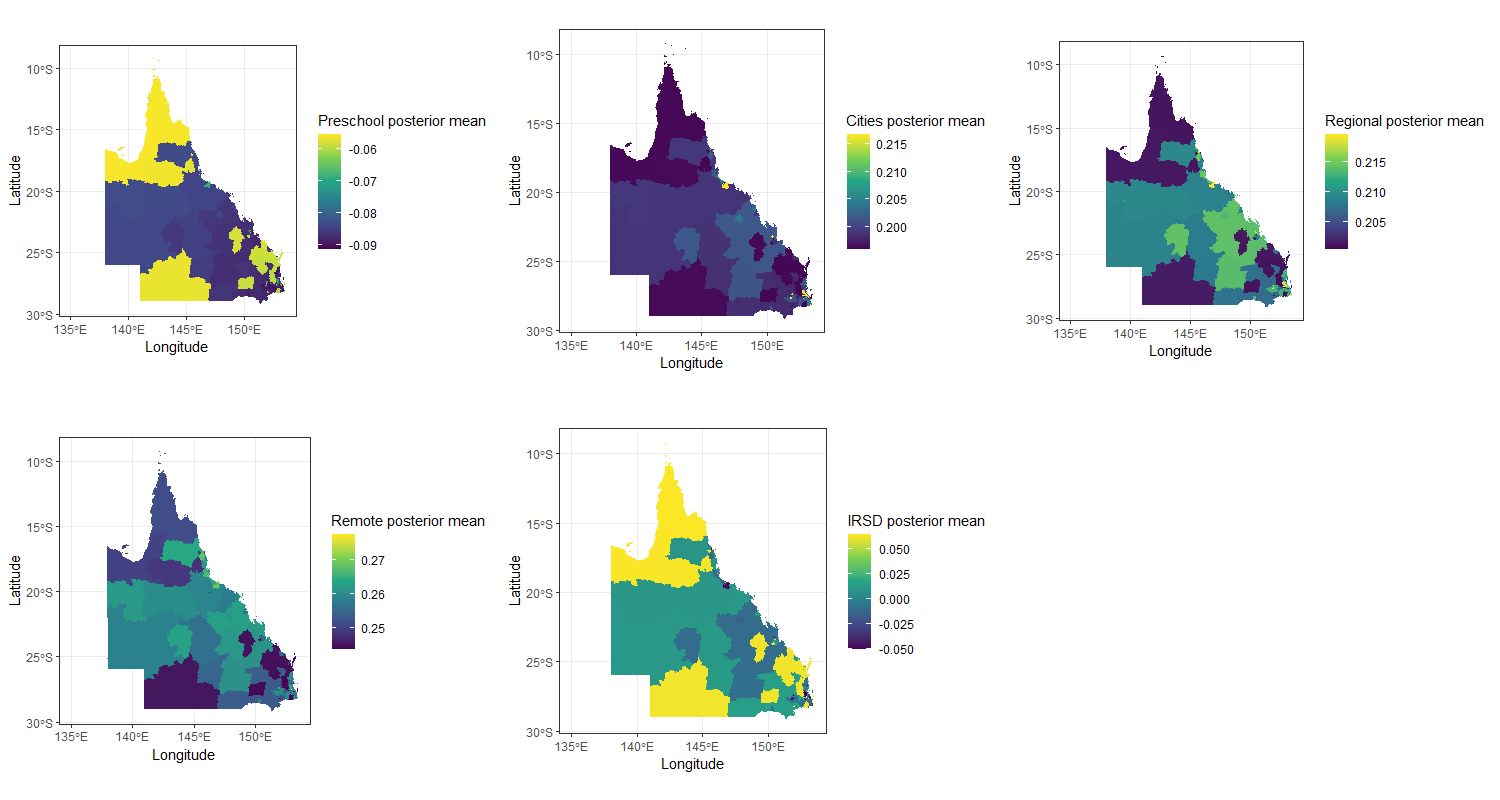}
     \caption{The spatial distribution of the posterior mean parameters derived from the proposed model.}
     \label{fig:enter-labela}
 \end{figure}

Figure \ref{fig:enter-labela} offers a geographical representation of five posterior mean parameters plotted on a map of Queensland. These values have been obtained through the proposed method, revealing that the relationship between the response variable (Vuln 1) and the covariates varies across different locations. 
To find the suitable structure of spatial weighs kernel for SDPMM for the case study, we used the WAIC. The WAIC values of the uniform and exponentially weighted kernels associated with different bandwidth priors can be found in Table \ref{tab:my-table6}. Comparison of the WAIC value leads to the conclusion that the exponential kernel is the most suitable for the real dataset. Additionally, Table \ref{mytab4} shows that as the complexity of the model (in terms of the number of knots) increases, the fit of the model to the data (WAIC) also improves. However, since the total number of SA2 is just 526, in this case study we assumed the number of knots to be 9.

\begin{table}[ht!]
\centering
\caption{Sensitivity analysis for the proposed algorithm with different bandwidth priors and kernel functions with 9 knots.}
\begin{tabular}{l|l|l}
\hline
Kernel type & Bandwidth Prior & WAIC \\
\hline
Uniform & $\varepsilon_{1i}, \varepsilon_{2i} \equiv \lambda$ & 3564.28 \\

 & $\varepsilon_{1i}, \varepsilon_{2i} \sim \text{Inverse Gamma}(1.5, \lambda^2/2)$ & 3568.94 \\
\hline
Exponential & $\varepsilon_{1i}, \varepsilon_{2i} \equiv \frac{\lambda^2}{2}$ & 3550.91 \\

 & $\varepsilon_{1i}, \varepsilon_{2i} \sim \text{Inverse Gamma}(1.5, \lambda^2/2)$ & 3565.42 \\
\hline
\end{tabular}
\label{tab:my-table6}
\end{table}

\begin{figure}[ht!]
    \centering
    \includegraphics[scale=0.5]{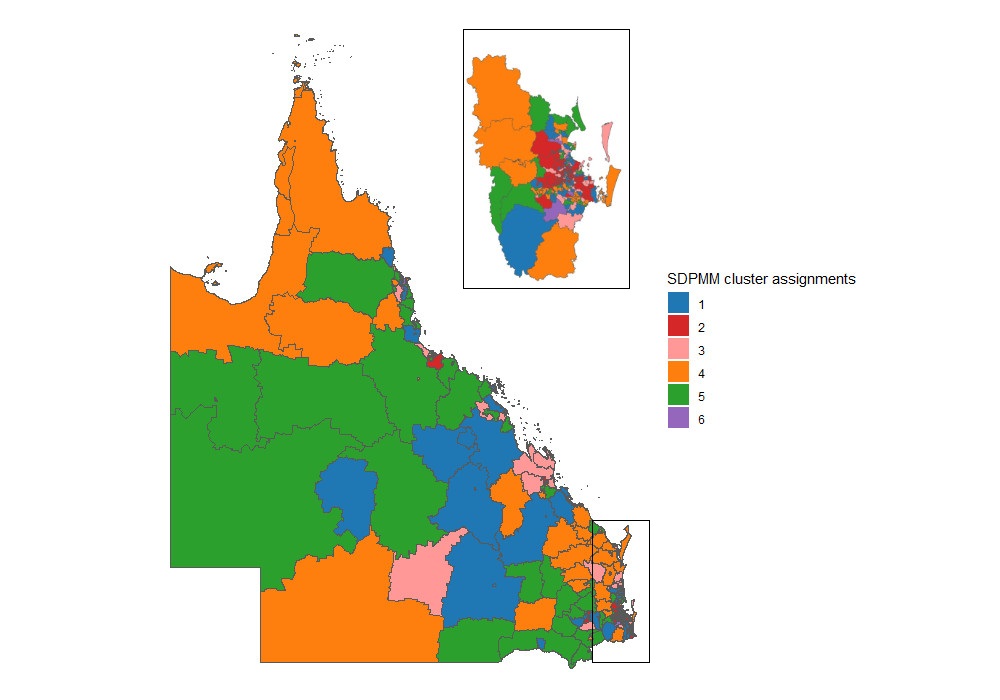}
    \caption{Cluster distribution from the proposed algorithm for the case study.}
    \label{fig:enter-labeqql}
\end{figure}

\begin{table}[ht!]
\caption{Sensitivity analysis for the number of knots in the spatial stick-breaking.}
\centering
\begin{tabular}{cc}
\toprule
Number of knots & WAIC \\
\midrule
9    &    3550.91 \\
19    &  2705.90  \\
24    &  1679.01   \\
32    & 1279.32 \\
\bottomrule
\end{tabular}
\label{mytab4}
\end{table}
Figure \ref{fig:enter-labeqql} showed the cluster distribution on the map obtained from the proposed algorithm with 6 clusters using Dahl's method. The cluster sizes are 124, 103, 90, 101 , 101 and 7. The strength of the proposed algorithm lies in its capability to create smaller cluster sizes compared to other cluster algorithms. This is beneficial for policy interventions targeting specific regions in Queensland, especially for identifying regions with high developmental vulnerabilities. Further, we provide a summary (Table \ref{table321}) for each of these clusters according to the parameter estimation and 95\% highest posterior density (HPD) interval.

Cluster 1 (124 out of 526) stands out due to its negative effect on the regression parameters for "Attendance at Preschool" with a narrow credible interval. The positive effects for the three levels of "Remoteness" are reliable, with the broadest uncertainty observed for the "Cities" parameters in comparison with the rest of the clusters. There's also some uncertainty in the "IRSD" parameters, which exhibit a negative effect, although they remain influential. Cluster 2 (103 out of 526) is characterized by a significant negative effect for "Attendance at Preschool" with a narrowest credible interval across the six clusters, indicating a strong impact and high confidence. Additionally, There are more positive effects for the "Cities", "Regional" and "Remote" parameters compared to Cluster 1, there is a more negative relationship for "IRSD" parameters compared to Cluster 1. Cluster 3 (90 out of 526) also exhibits a significant negative effect for "Attendance at Preschool" parameters but with a broader credible interval, indicating a strong impact with more uncertainty. The positive effects for "Remoteness" parameters are still significant and confident, with the broadest uncertainty for the "Regional" parameters across the six clusters, "IRSD" exhibits the most negative parameters  in this cluster in comparison with the rest. Cluster 4 (101 out of 526) maintains a significant negative effect for "Attendance at Preschool" with a narrow credible interval, with a  positive effects for "Remoteness" parameters. In this cluster "IRSD" has a positive effect with a narrow credible interval in comparison with clusters 1, 2 and 3. Cluster 5 (101 out of 526) has a negative effect for "Attendance at Preschool" and a narrow credible interval. Similar to Cluster 4, the positive effects for "Remoteness". But in this cluster "IRSD" has more positive effect in comparison with cluster 4. Cluster 6 (7 out of 526) stands out with it negative effect for "Attendance at Preschool" even though it has a wider credible interval. The positive effects for "Remoteness" are similar to the previous clusters, with the narrowest credible intervals for the "Cities, Regional, and Remote" parameters, also the "IRSD" has a negative effect with a narrow credible interval. 

These clusters are distinguished primarily by the magnitude and certainty of the effect of "Attendance at Preschool" and the reliability of the "Remoteness" and "IRSD". Cluster 1, 4,5, and 6 share a strong negative impact of "Attendance at Preschool" with narrow credible intervals. Cluster 2 , with a broader credible interval, indicates more uncertainty in the impact of "Attendance at Preschool". Cluster 3, despite having a wider credible interval for "Attendance at Preschool" still shows a significant negative effect. Additionally, the "IRSD" exhibits variations across clusters, adding another layer of distinction.

\begin{table}[ht!]
\caption{Parameter estimates and their 95\% highest posterior density (HPD) intervals for the six clusters identified.  $\hat{\beta_1}$: attendance at preschool parameters, $\hat{\beta_2}, \hat{\beta_3}, \hat{\beta_4}$: remoteness parameters, with three levels (cities, regional, and remote), and $\hat{\beta_5}$: IRSD parameters.  }
\centering
\scalebox{0.8}{
\begin{tabular}{c|c|c|c|c|c}
\toprule
Cluster & $\hat{\beta}_1$ & $\hat{\beta}_2$ & $\hat{\beta}_3$ & $\hat{\beta}_4$ & $\hat{\beta}_5$ \\
\midrule
1 & -0.083 (-0.126, -0.003) & 0.206 (0.199, 0.216) & 0.214 (0.197, 0.235) & 0.259 (0.221, 0.293) & -0.014 (-0.029, -0.001) \\
2 & -0.075 (-0.112, -0.035) & 0.216 (0.205, 0.230) & 0.217 (0.200, 0.232) & 0.262 (0.231, 0.294) & -0.049 (-0.070,  -0.031) \\
3 & -0.095 (-0.138, -0.046) & 0.198 (0.189, 0.207)& 0.213 (0.199, 0.231) & 0.259 (0.221, 0.299) & -0.005 (-0.018, 0.009) \\
4 & -0.056 (-0.112, -0.004) & 0.196 (0.186, 0.205)& 0.202 (0.178, 0.221) & 0.244 (0.221, 0.278)& 0.063 (0.044, 0.086)\\
5 & -0.087 (-0.134, -0.040) & 0.199 (0.189, 0.207) & 0.207 (0.187, 0.224) & 0.255 (0.224, 0.296) & 0.013 (-0.003, 0.036) \\
6 &-0.125 (-0.173, -0.102) & 0.200 (0.194, 0.208) &0.192 (0.177, 0.199) &0.256 (0.235, 0.278) & -0.017 (-0.028, -0.010) \\
\bottomrule
\end{tabular}}

\label{table321}
\end{table}

\section*{Discussion}

In this paper, we introduce a new statistical framework aimed at addressing the challenges in clustering posed by spatially varying relationships within regression analysis. Specifically, we present a Bayesian model that integrates geographically weighted regression with a spatial Dirichlet process to cluster relevant model parameters. This solution therefore not only identifies clusters of the model parameters but effectively captures the inherent heterogeneity present in spatial data. Our exploration encompasses various weighting schemes designed to effectively model the complex spatial interaction between neighborhood characteristics and the positioning of key points (or "knots"). This modelling is supported by a discussion of Bayesian model selection criteria, a crucial step in the analysis process that ensures selection of an appropriate and well-fitted model. Spatial variation in the effects of covariates empowers our model to provide a better fit to spatial data compared to conventional models, offering insights into the complex patterns of heterogeneity across diverse geographical locations. Additionally, making smaller group sizes helps decision-makers identify which regions need more help. To demonstrate the efficacy of our methodology, we have presented a simulation study. Moreover, we have extended our investigation to a real-world application: a thorough analysis of the factors influencing children's development indicators in Queensland. Through this practical example, we showcase the benefits of our proposed approach, emphasizing its ability to find hidden dynamics that might otherwise remain obscured.

In our case study, we aimed to explore the influential factors affecting child development vulnerability in Queensland's statistical area level 2 (SA2) regions. Our analysis utilised a dataset consisting of 526 observations, each corresponding to one of Queensland's SA2 regions. The dataset included various explanatory variables, including preschool attendance, remoteness factors, and socio-economic factors. The primary objective was to identify spatial clusters of children's vulnerability and gain insights into the regional disparities in children's development domains. To select the appropriate spatial weights kernel for our model, we employed WAIC and found the uniform kernel, suggesting that it provides a better fit for our real dataset. Furthermore, we performed a sensitivity analysis to determine the optimal number of knots in the spatial stick-breaking process and found increasing the complexity of the model by adding more knots improved its fit to the data. We selected 9 knots as the optimal number for our analysis. Using the selected model with an exponential kernel and 10 knots, we applied the proposed algorithm to identify spatial clusters of child vulnerability. Our analysis revealed a total of 6 clusters across Queensland's SA2 regions. These clusters vary in size, with the largest containing 124 regions and the smallest comprising only 7 regions. The ability of the proposed algorithm to create smaller cluster sizes is noteworthy, as it allows for more targeted policy interventions in regions with specific developmental needs. Moreover, 
we conducted a detailed analysis of the clusters to understand their characteristics and implications. For instance, the presence of smaller clusters may indicate isolated areas with unique developmental challenges that require tailored interventions. In contrast, larger clusters could represent regions with similar vulnerabilities, suggesting the need for broader policy strategies. These findings offer valuable insights for policymakers and stakeholders interested in addressing child development disparities in Queensland.

While our paper represents a step forward in the field of spatial regression, it is essential to acknowledge the avenues for further exploration that our research did not study. For instance, while we thoroughly examined the full model that incorporates all relevant covariates, we did not delve into methodologies for variable selection within the context of clustered regression. This presents a clear direction for future research, where innovative approaches for selecting the most influential variables among a clustered regression could enhance the performance of our model even further. Additionally, our work touched upon the utilisation of the spatial Dirichlet process mixture model (SDPMM) to derive cluster information for regression coefficients. However, we acknowledge that the posterior distribution of the cluster count might not always be accurately estimated through the SDPMM, as demonstrated by Miller and Harrison \cite{miller2013simple}. Our simulation studies confirm this observation. This area emerges as a critical focus for future studies. Another area for future work involves expanding our methodology to accommodate non-Gaussian data distributions, a direction that holds promise for a wider range of applications. Moreover, the pursuit of adapting our model to handle multivariate response scenarios represents an essential avenue for future exploration, offering the potential to unlock insights and applications across various domains.

\section*{Methods}

This section outlines the proposed model called the spatial Dirichlet process clustered heterogeneous regression model. The model utilises a non-parametric spatial Dirichlet mixture model applied to the regression coefficients of the geographically weighted regression model. The model is cast in a Bayesian framework.

 \subsection*{Bayesian Geographical Weighted Regression} \label{weight}
 The Bayesian geographically weighted regression (BGWR) model can be described as follows. Given diagonal weight matrix \( W(s) \) for a location $s$, the likelihood for each \( y(s) \) is:
\begin{equation}
\centering
y(s)|\beta(s), x(s), W_i(s), \sigma^2(s) \sim \mathcal{N}(x^T(s)\beta(s), \sigma^2(s) W_i^{-1}(s))
\end{equation}

\noindent where \( y(s) \) is the \(i\)th observation of the dependent variable, \( x(s) \) is the \(i\)th row (or observation) from the design matrix \( X \) and \( W_i(s) \) is the \(i\)th diagonal element from the spatial weight matrix \( W(s) \).  The weighted matrix $W(s)$ is constructed to identify the relative influence of neighbouring regions on the parameter estimates at locations.
\\
When working with areal data, the graph distance is an alternative distance metric that can be used. It is based on the concept of a graph, where $V=\{v_1,...,v_m\}$ represents the set of nodes (locations) and $E(G)=\{e_1,...,e_n\}$ represents the set of edges connecting these nodes. The graph distance is defined as the distance between any two nodes in the graph \cite{ma2020heterogeneous}.

\[ W(s)=\begin{cases} 
      |V(e)| &   \text{if $e$ is the shortest distance connecting a pair of nodes,}\\
      \infty & \text{if the two nodes are not connected}
   \end{cases}
\]
 where $|V(e)|$ is the number of edges in $e$  \cite{gao2010survey}. The graph distance-based weighted function is given as:
\[ W(s)=\begin{cases} 
      1 &   \text{if $d_i(s) \leq b$,}\\
      f(d_i(s)^ b) & \text{ otherwise}
   \end{cases}
\]
where $d_i(s)^ b$ is the graph distance between locations $i$ and $s$, $f$ is a weighting function, and $b$ represents the bandwidth. In this study, we suppose that $f()$ is a negative exponential function \cite{ma2021geographically}, so,
\[ W(s)=\begin{cases} \label{graph}
      1 &   \text{if $d_i(s) \leq 1$,}\\
      e^{(-d_i(s)/b)} & \text{ otherwise}
   \end{cases}
\]
 where $b$ represents the bandwidth that controls the decay with respect to distance \cite{cho2010geographically}. Here, $d_i(s)$ indicates that an observation far away from the location of interest contributes little to the estimate of parameters at that location. In this paper, we used the graph distance and the greater circle distances \cite{bullock2007great} and both of these methods show consistent parameters. 
 The proposed model is constructed in a Bayesian framework with conjugate priors on the regression parameters and other model terms. The full model is given in the (Full Bayesian spatial Dirichlet process mixture prior cluster heterogeneous regression) section.
\subsection*{Heterogeneous Regression with Spatial Dirichlet Process Mixture Prior}
In a Bayesian framework, coefficient clustering can be achieved by using a Dirichlet process mixture model (DPMM). This approach links the response variable to the covariates through cluster membership. The DPMM is defined by a probability measure $G$ that follows a Dirichlet process, denoted as $G \sim (\alpha, G_0)$, where $\alpha$ is the concentration parameter and $G_0$ is the base distribution \cite{ma2020heterogeneous}. Hence,

\begin{equation}
    (G(A_1),...,G(A_k)) \sim Dirichlet (\alpha G_0(A_1),...,\alpha G_0(A_k))
\end{equation}
where $(A_1,...,A_k)$ is a finite measurable partition of the space $\Omega$, and the variable $k$ represents the number of components or clusters in a (DPMM). 
Several formulas have been proposed in the literature for specifying the DPMM's parameters and incorporating spatial dependencies \cite{quintana2022dependent, yamato2020dirichlet}. A popular approach is the spatial stick-breaking algorithm \cite{reich2007multivariate,sethuraman1994constructive}, which in a BGWR setup is applied at each location as follows:
\begin{equation}
\begin{aligned}
G(s) &= \sum_{i=1}^{K} p_i(s) \delta(\theta_i) \\
p_1(s) &= V_1(s) \\
p_i(s) &= V_i(s) \prod_{j=1}^{i-1} (1-V_j(s)) \quad \text{for all } i>1 \\
V_i(s) &= l_i(s) V_i \\
V_i &\sim \text{Beta}(a, b)
\end{aligned}
\end{equation}

\noindent where $\delta(\theta_i)$ is the Dirac distribution with a point mass at ($\theta_i)$, and $p_i (s)$ is a random probability weight between 0 and 1. The distribution of $G(s)$ depends on $V_i$ and $\theta_i$; the distribution varies according to the kernel function $l_i(s)$. However, the spatial distributions of the kernel function $l_i(s)$ vary, constrained within the interval $[0, 1]$. These functions are centered at knots $\psi_i=(\psi_{1i}, \psi_{2i})$, and the degree of spread is determined by the bandwidth parameter $\epsilon_i=(\epsilon_{1i}, \epsilon_{2i})$. Both the knots and bandwidths are treated as unknown parameters with independent prior distributions, unrelated to $V_i$ and $\theta_i$. The knots $\psi_i$ are assigned independent uniform priors, covering the bounded spatial domain. The bandwidths can be modelled to be uniform for all kernel functions or can vary across kernel functions, following specified prior distributions \cite{hosseinpouri2019area,reich2007multivariate}. The most common kernels are the uniform and the square exponential functions. This kernel can take different formats. Table \ref{table1} provides examples of the most popular kernels used for the spatial stick-breaking configuration.
 

\begin{table}[htbp]
\centering
\caption{Examples of kernel functions, where $IG$ presents inverse Gamma function.}
\label{table_true}
\begin{tabular}{|p{3.5cm}|p{4.5cm}|p{4cm}|}
\hline
Name & $l_i(s)$ & Model for $\varepsilon_{1i}$ and $\varepsilon_{2i}$  \\
\hline
Uniform & $\prod_{j=1}^{2} I\left(|s_j - \psi_{ji}| < \frac{\varepsilon_{ji}}{2}\right)$ & $\varepsilon_{1i}, \varepsilon_{2i} \equiv \lambda$  \\
\hline
Uniform & $\prod_{j=1}^{2} I\left(|s_j - \psi_{ji}| < \frac{\varepsilon_{ji}}{2}\right)$ & $\varepsilon_{1i}, \varepsilon_{2i} \sim \text{Exp}(\lambda)$  \\
\hline
Squared exp. & $\prod_{j=1}^{2} \exp\left(-\frac{(s_j - \psi_{ji})^2}{2\varepsilon_{ji}^2}\right)$ & $\varepsilon_{1i}, \varepsilon_{2i} \equiv \frac{\lambda^2}{2}$ \\
\hline
Squared exp. & $\prod_{j=1}^{2} \exp\left(-\frac{(s_j - \psi_{ji})^2}{2\varepsilon_{ji}^2}\right)$ & $\varepsilon_{1i}, \varepsilon_{2i} \sim \text{IG}(1.5, \lambda^2/2)$  \\
\hline
\end{tabular}
\label{table1}
\end{table}

A vector of latent allocation variables $Z$ is generated to characterize the clustering explicitly. Let $Z_{n,k}=\{z_1.,...,z_n\}$, where $z_i \in \{1,...,k\}$ and $1\leq i \leq n$ represents all possible clustering of $n$ observations into $K$ clusters.
\subsection*{Full Bayesian Spatial Dirichlet Process Mixture Prior Cluster Heterogeneous Regression } \label{model}
Adapting the spatial Dirichlet process to the heterogeneous regression model, we focus on clustering of spatial coefficients $\beta(s_1),...,\beta(s_n)$ and $\beta(s_i)=\beta_{{z_i}} \in \{\beta_1,...,\beta_k\}$. The full model is described as follows with the most commonly adopted priors:
\begin{equation}
\label{RJJ}
y(s)|\beta(s), x(s), W_i(s), \sigma^2(s) \sim \mathcal{N}(  x^T(s)\beta(s_{z_{i}}), \sigma^2(s) W_i^{-1}(s))
\end{equation}

\begin{equation}
W_i(s)=f(d_i|b)
\end{equation}

\begin{equation}
b \sim Uniform(0,D)
\end{equation}

\begin{equation}
\label{eq:beta_prior}
\beta_{z_i} \sim \mathcal{N}_p(\mu_{z_i}, \Sigma_{z_i})
\end{equation}

\begin{equation}
z_i\sim categorical(p_1(s),p_2(s),...,p_k(s))
\end{equation}

\begin{equation}
\mu_k | \Sigma_k \sim \mathcal{N}_p(m_k,  \Sigma_k)
\end{equation}

\begin{equation}
\Sigma_k \sim IW(D_k,c_k)
\end{equation}

\begin{equation}
\sigma^2(s) \sim IGamma(\alpha_1,\alpha_2)
\end{equation}

\begin{equation}
P(z_i = k | p) = p_k(s)
\end{equation}

\begin{equation}
p_1(s)=V_1(s),
p_k(s)=V_k(s) \prod _{j=1}^{K-1} (1-V_j(s) ),
V_k(s)=l_k(s) V_k   
\end{equation}

\begin{equation}
V_k \sim Beta(a_v,b_v)
\end{equation}

Here, the response variable \(Y\) is assumed to follow a Gaussian distribution; the design matrix representing the predictors is denoted by $X$, and  the spatial weight matrix $W(s)$ depends on two key aspects: the distance between observations, represented as \(d_i\), and a parameter \(b\), which controls the bandwidth. This bandwidth is assumed to follow a uniform distribution between 0 and a certain value \(D\). $f$ is the graph weighting function. The regression coefficients \(\beta_{z_i}\) are associated with a specific group, or cluster, \(z_i\), for a particular observation. The mean and spread of cluster \(z_i\) are denoted as \(\mu\) and \(\Sigma_{z_i}\) ,respectively, and the maximum number of possible clusters is \(K\).

The hyper-parameter \(m_k\) is a prior mean value for the \(\mu_{z_i}\) and   \(\Sigma_k\) is a way to express how different a cluster can be. Similarly, \(D_k\) is the scale matrix, and $c_k >p-1$ is the degrees of freedom. Another important aspect is the variation in the data, which is \(\sigma^2(s)\). This variation changes across locations and follows a specific prior pattern, which is assumed to be an inverse Gamma distribution with parameters \(\alpha_1\) and \(\alpha_2\).

 we focus on the probability \(P(z_i = k)\) that observation \(i\) belongs to cluster \(k\). This assignment probability at a specific location \(s\) is denoted as \(p_k(s)\). For the clusters, we also consider "stick-breaking weights", denoted by \(V_k(s)\), which change across locations. The values \(a_v\) and \(b_v\) are related to how these weights are determined using a beta distribution. Here $\sum_{j=1}^{k} p_k (s) = 1$ almost surely under the constraint that $V_k (s) = 1$ for all locations $s$ \cite{ishwaran2001gibbs}.

\section*{Bayesian Estimation and Inference}
This section covers using MCMC to obtain samples from posterior distributions of model parameters. It explains the sampling scheme, covers the use of posterior inference for cluster assignments, and methods for evaluating accuracy.
\subsection*{The MCMC sampling schemes}
The main R function for the model is implemented using the nimble package \cite{de2017programming}. This function encapsulates
the model and provides an interface for executing the MCMC sampling scheme, performing
posterior inference, and evaluating estimation performance and clustering accuracy. The model
itself is wrapped within a nimbleCode function, which allows the nimble package to generate and
compile C++ code to execute the MCMC sampling scheme efficiently. This can result in substantial
speed improvements over pure R implementations, especially for models with large datasets or
complex parameter space.
In the context of the proposed algorithm, the nimble package provides several MCMC sampling
methods, including the popular Gibbs and Metropolis-Hasting algorithms for inferring the posterior distribution of the regression and other model parameters. Nimble  also allows for the specification of priors and likelihood functions for the parameters to customise the MCMC sampling process. In our study, the Gibbs
sampling algorithm was used to obtain the clusters of the parameters. 

Block Gibbs sampling is a MCMC technique used for sampling from the joint distribution of multiple random variables. The primary idea behind block sampling is to group related variables together into "blocks" and sample them jointly, which can improve the efficiency and convergence of the sampling process \cite{yu2010document}. An explanation of this sampling algorithm for the proposed algorithm can be found in the Appendix.
\subsection*{Cluster configurations} \label{conf}
Two methods are used to determine cluster configurations. In the first, the estimated parameters, together with the cluster assignments $Z_{n,k}$ are determined for each replicate from the best post-burn-in iteration selected using Dahl’s method \cite{dahl2006model}, which involves estimating the clustering of observations through a least-squares model-based approach that draws from the posterior distribution. In this method, membership matrices for each iteration, denoted as \(B^{(1)}, \ldots, B^{(M)}\), with  \(M\) being the number of post-burn-in MCMC iterations, are computed. The membership matrix for the \(c\)-th iteration, \(B^{(c)}\) is defined as:
\begin{equation}
B^{(c)} = (B^{(c)}(i, j))_{i,j \in \{1:n\}} = \mathbf{1}(z^{(c)}_i = z^{(c)}_j)_{n \times n}
\end{equation}

\noindent where $\mathbf{1}(\cdot)$ represents the indicator function. The entries $B^{(c)}(i, j)$ take values in $\{0, 1\}$ for all $i, j = 1, \ldots, n$ and $c = 1, \ldots, M$. When $B^{(c)}(i, j) = 1$, it indicates that observations $i$, and $j$ belong to the same cluster in the $c$th iteration.\\
To obtain an empirical estimate of the probability for locations $i$ and $j$ being in the same cluster, the average of $B^{(1)}, \ldots, B^{(M)}$ can be calculated as:
\begin{equation}
\overline{B} = \frac{1}{M} \sum_{c=1}^M B^{(c)}
\end{equation}

\noindent where $\sum$ denotes the element-wise summation of matrices. The $(i, j)$th entry of $\overline{B}$ provides this empirical estimate.\\
Subsequently, the iteration that exhibits the least squared distance to $\overline{B}$ is determined as:
\begin{equation}
C_{LS} = \arg \min_{c \in \{1:M\}} \left[\sum_{i=1}^n \sum_{j=1}^n (B^{(c)}(i, j) - \overline{B}(i, j))^2\right]
\end{equation}
where $B^{(c)}(i, j)$ represents the $(i, j)$th entry of $B(c)$, and $\overline{B}(i, j)$ denotes the $(i, j)$th entry of $\overline{B}$. The least-squares clustering offers an advantage in that it leverages information from all clusterings through the empirical pairwise probability matrix $\overline{B}$. \\

The second method utilised here is the posterior mode method. This method leverages posterior samples
from iterations associated with $z_i$, where $z$ denotes the cluster assignments specific to each
region. Each iteration generates a new set of cluster assignments $z$, which are dependent on
the parameters. Consequently, following multiple iterations, each region will have an empirical posterior distribution of
cluster assignments $z$. The mode indicates the cluster with the highest probability of assignment
for a given region. 
\subsection*{Cluster Accuracy}
In order to assess the accuracy of the proposed algorithm, we compared the cluster configurations with the true labels provided for the simulated data. It is important to note that while the true labels are available for the simulated data, such information is not readily available for real-world datasets. In practice, true labels are often unknown, which poses a challenge for the evaluation of clustering accuracy.
In this study, we utilised the Rand index (RI) \cite{rand1971objective}. This index measures the level of similarities between two sets of cluster assignments, labelled as \(C\) and \(C'\), with respect to a given dataset \(X = \{x_1, x_2, \ldots, x_n\}\). Each data point  \(x(s)\) is assigned a cluster label \(c_i\) in  \(C\) and \(c'_i\) in \(C'\).
The RI is computed using the following formula:
\begin{equation}
\text{RI} = \frac{a + b}{a + b + c + d}.
\end{equation}

\noindent Hhere \(a\), represents the number of pairs of data points that are in the same cluster in both \(C\) and \(C'\) (true positives); \(b\) indicates the number of pairs of data points that are in different clusters in both \(C\) and \(C'\) (true negatives); \(c\) represents the number of pairs of data points that are in the same cluster in \(C\) but in different clusters in \(C'\) (false positives); and \(d\) stands for the number of pairs of data points that are in different clusters \(C\) but in the same cluster in \(C'\) (false negatives).

The Rand index ranges from 0 to 1, with a value of 1 denoting a complete concordance between the two clusterings (both $C$ and \(C'\)  perfectly agree on all pairs of data points). Conversely, a value close to 0 indicates a weak level of agreement between the two clusterings.

\section*{Conclusion}

This paper introduces a method called the spatial Dirichlet process clustered heterogeneous regression model. The method employs a non-parametric Bayesian clustering approach to group the spatially varying regression parameters of a Bayesian geographically weighted regression, and also determines the best number and arrangement of clusters. The model uses advanced Bayesian techniques to cluster the parameters and determine the best number and arrangement of clusters. The model's abilities were demonstrated using simulated data and then applied to actual data related to children's development vulnerabilities in their first year of school. In this application, the model successfully identified key factors. This approach enhances our understanding of how children develop in various regions, revealing the factors that impact their health and well-being. With these insights, policymakers can create targeted policies that are suited to each area's unique characteristics. As a result, this innovative method not only improves the suite of analytical tools but also contributes to the broader goal of enhancing the health and development prospects of children in different places.
\bibliography{sample}
\section*{Supporting information}
\textbf{S1 Appendix}. Blocked MCMC for a Spatial Dirichlet Process Mixture Model
(SDPMM)
\section*{Acknowledgements}
We would like to express our gratitude to the team at Children's Health Queensland and the
Centre for Data Science for their invaluable assistance and support in this project.

\section*{Data Availability Statement}
All the datasets used in this article are publicly accessible and free to download. Anyone interested can access them without special privileges. Likewise, the authors did not have any special privileges when accessing the data for analysis in this article. The datasets can be obtained from the following sources: Children’s Health Data is sourced from the Australian Early Development Census, available upon request at \url{https://www.aedc.gov.au/data-explorer/}. The Explanatory Data is obtained from the Australian Bureau of Statistics and is publicly available at \url{https://www.abs.gov.au/census/find-census-data/quickstats/2021/3}.

\section*{Competing interests}
The authors have declared that no competing interests exist.

\section*{Author Contributions}
Conceptualization: Wala Draidi Areed, Aiden Price, Helen Thompson, Kerrie Mengersen.
Data curation: Wala Draidi Areed, Aiden Price, Reid Malseed, Kerrie Mengersen.
Formal analysis: Wala Draidi Areed.
Investigation: Wala Draidi Areed.
Methodology: Wala Draidi Areed, Kerrie Mengersen.
Software: Wala Draidi Areed.
Supervision: Aiden Price, Helen Thompson, Reid Malseed, Kerrie
Mengersen.
Validation: Wala Draidi Areed.
Visualization: Wala Draidi Areed.
Writing – original draft: Wala Draidi Areed.
Writing – review  and editing: Wala Draidi Areed, Aiden Price,  Helen Thompson,
Reid Malseed, Kerrie Mengersen.

\newpage
\begin{appendices}
\section*{S1. Appendix. Blocked MCMC for a Spatial Dirichlet Process Mixture Model (SDPMM)}
In this section, we use a blocked sampler. This method updates groups of parameters all at once to make computations more efficient. By updating these parameter blocks together, we can achieve faster convergence and enhance the overall performance of the chain.\\
Given the model specifications, the joint distribution (prior to normalization) can be expressed as:
\[
p(y(s), {Z}, {\beta(s)}, {\mu}, {\Sigma}, \sigma^2(s), {V}, b) = p({y(s)}|{\beta(s)}, {Z}, \sigma^2(s)) \times p({Z}|{V}) \times p({\beta(s)}|{\mu}, {\Sigma}, {Z}) \times p({\mu}|{\Sigma}) \times p({\Sigma}) \times p(\sigma^2(s)) \times p({V}) \times p(b) 
\]
\textbf{Block Gibbs Sampling}
\begin{itemize}
    \item \textbf{Initialization}

Choose initial values for \( \beta(s), \mu, \Sigma, \sigma^2(s), V, b \) and latent variables \( Z \).

\item \textbf{Gibbs Sampling Iterations}

A. Sample \(Z\) given other parameters and data:

For each observation \(i\), sample \(Z_i\) from its categorical distribution based on likelihood and prior terms:
\[
p(Z_i = k|{y(s)}, {\beta(s)}, {\mu}, {\Sigma}, \sigma^2, {V}, b) \propto p(y(s)|\beta(s_{z_{i}}), x(s), \sigma^2(s)) \times p(Z_i = k|{V}) 
\]

B. Sample other blocks of parameters:

Sample \( \beta(s) \) given \( Z, \mu, \Sigma, y(s), x(s) \) and other relevant variables: 
\[
p({\beta(s)}|{y(s)}, {Z}, {\mu}, {\Sigma}, \sigma^2(s), {V}, b) \propto p({y(s)}|{Z}, {\beta(s)}, \sigma^2(s)) \times p({\beta(s)}|{\mu}_Z, {\Sigma}_Z)
\]

C.  Compute the gradient of the log posterior with respect to \( \beta(s) \) using the derived gradients from the likelihood and prior terms:
\[
\frac{\partial}{\partial \beta(s)} \log p(\beta(s), Z, \mu, \Sigma, \sigma^2(s), V, b | y(s)) = \sum_i 2x(s)(y(s) - x^T (s)\beta(s_{z_{i}}))
\]

D. Update \( \beta(s) \) using the gradient information and suitable sampling algorithm (e.g., Metropolis-Hastings).

E. Sampling \( \mu \) and \( \Sigma \):
   
    Given the conjugacy in the model, \( \mu_k \) and \( \Sigma_k \) can be jointly sampled from their respective full conditional distributions.
   
F.  For each cluster \( k \), compute the posterior distribution of \( \mu_k \) and \( \Sigma_k \) using the corresponding data and prior:
   
\[
p(\mu_k, \Sigma_k | \text{data and prior information}) \propto p(\beta(s_{z_{i}}) | \mu_k, \Sigma_k) \times p(\mu_k | \Sigma_k) \times p(\Sigma_k)
\]

G.  Update \( \mu \) and \( \Sigma \) using the sampled values from their posterior distributions.

H. Sample \( \sigma^2(s) \) given \( y(s), x, \beta(s), Z \), and other variables:

    For each spatial location \( s \), sample \( \sigma^2(s) \) using the inverse-gamma distribution:
\[
\sigma^2(s) \sim \text{Inverse-Gamma}(\alpha_1 , \alpha_2 )
\]

I. Sample \( V \) and \( b \) given their priors and their relationship to \( Z \):

  $ V_k$ represents the parameter associated with latent variable assignment for cluster $k$. It determines the probabilities of assigning observations to cluster $k$. This step updates the assignment probabilities based on the data and relationships between $V$ and the latent variables $Z$. 
  
  Sample \( V_k \) using the beta distribution with updated parameters based on the data and relationship to \( Z \):
   
\[
V_k \sim \text{Beta}(a_v , b_v )
\]
    
\end{itemize}

\end{appendices}
\end{document}